\documentclass[12pt,aps,prc,showpacs,amsmath,showkeys]{revtex4}

\usepackage[dvips]{graphicx}
\usepackage{epstopdf}
\usepackage{amssymb}
\usepackage{epstopdf}
\newcommand{\ba}{\begin{eqnarray}}
\newcommand{\ea}{\end{eqnarray}}
\newcommand{\bmath}{\begin{mathletters}}
\newcommand{\emath}{\end{mathletters}}
\newcommand{\ban}{\begin{eqnarray*}}
\newcommand{\ean}{\end{eqnarray*}}

\draft

\begin{document}

\title{Pseudospin symmetry and effective field theory}
\author{Joseph N. Ginocchio}
\address{MS B283, Theoretical Division, Los Alamos National Laboratory, Los
Alamos, New Mexico 87545, USA}
\date{\today}

\begin{abstract}
Pseudopsin has been shown to be approximately conserved in nuclei. We investigate whether or not there  is an advantage in using the pseudospin operators as well as the spin operators in the description of the nucleon-nucleon interaction in effective field theory. We conclude that, indeed, there is an advantage.
\end {abstract}

\keywords{ Symmetry,  Pseudospin, Effective field theory, Chiral perturbation theory, Relativistic mean field theory, Spin}

\pacs{ 21.60.-n, 21.10.-k, 02.20.-a}

\maketitle

\newpage
Pseudospin symmetry is a relativistic symmetry of the Dirac Hamiltonian that occurs when the scalar and vector potentials are equal in magnitude and opposite in sign \cite  {bell}. This condition approximately holds for the relativistic mean fields of nuclei \cite {madland}. Indeed, nuclear energy levels and transition rates are consistent with approximate pseudospin symmetry  \cite {gino05}. However, the basic nuclear interaction between nucleons is expressed in terms of spin operators and not pseudospin operators. In particular effective field theory involves the expansion of the nucleon-nucleon interaction in powers of the momentum \cite {eft} and pseudospin involves the intertwining of spin and momentum. In this paper we would like to explore if the inclusion of pseudospin operators in the nucleon-nucleon interaction is useful and instructive. 

Pseudospin symmetry is an SU(2) symmetry as is spin symmetry. The relativistic generators for the pseudospin algebra,
${\tilde{S}}_i(i=x,y,z)$, are \cite{ami}
\begin{equation} 
{\tilde S}_i = \left(\begin{array}{cc} 
{\tilde s}_i & \quad 0 \\[2pt]
0 & \quad s_i\end{array}\right)= \left(\begin{array}{cc}
U_p\, {s}_iU_p & \quad 0 \\[2pt]
0 & \quad {s}_i
\end{array}\right)\,,
\label{psg}
\end{equation}
where ${ s}_i=\sigma_i/2$ are the usual spin generators, $\sigma_i$
the Pauli matrices, and $U_p={\sigma}\cdot {\hat p}$ is the
momentum-helicity unitary operator \cite{draayer},  and ${\hat p}_i= {{ p_i}\over p}$ is the unit three momentum of a single nucleon. The four by four nature of the generators results from the fact that they are  relativistic generators. The generators for the non-relativistic pseudospin algebra are
\begin{equation}
{\tilde s}_i = U_p\ {  s}_i\ U_p = \sigma \cdot  {\hat p}\ {\hat p}_i - s_i.
\label{gen1}
\end{equation}
We note that, although the pseudospin generators depend on momentum, they depend on the unit vector of momentum and therefore are equivalent to spin as far as momentum power counting in effective field theory.

These are the generators for a one particle system moving in a mean field. For a many nucleon system interacting, the pseudospin generators need to be translationally invariant. Therefore we use intrinsic momenta for the $A$ nucleons, $q_{i,n} = p_{i,n} - {\Sigma_{k=1}^{A} \ p_{i,k}\over A}$, with $ n = 1\dots A$,

%\begin{equation} 
%{\tilde S}_{i,n} = \left(\begin{array}{cc} 
%{\tilde s}_{i ,n}& \quad 0 \\[2pt]
%0 & \quad s_{i,n} \end{array}\right)= \left(\begin{array}{cc}
%U_{q_n}\, {s}_{i,n}U_{q_n} & \quad 0 \\[2pt]
%0 & \quad {s}_{i,n}
%\end{array}\right),
%\label{psg}
%\end{equation}

\begin{equation}
{\tilde s}_{i,n} = \sigma_n \cdot  {\hat {q_n}}\ {\hat {q}}_{i ,n}- s_{i,n}.
\label{genA}
\end{equation}

For the two nucleon interaction, $q_{i,1}= {p_{i,1} -  p_{i,2}\over 2} = - q_{i,2} = p$, and hence the generators are

\begin{equation}
{\tilde s}_{i,1} =  \sigma_1 \cdot  {\hat {p}}\ {\hat {p}}_{i }- s_{i1},\
{\tilde s}_{i,2} =  \sigma_2 \cdot  {\hat {p}}\ {\hat {p}}_{i }- s_{i,2}.
\label{gen}
\end{equation}

First we consider the pseudospin-pseudospin interaction
\begin{equation}
{\tilde s}_1\cdot {\tilde s}_2 = ( \sigma_1 \cdot  {\hat {p}}\ {\hat {p}}- s_{1})\cdot ( \sigma_2 \cdot  {\hat {p}}\ {\hat {p}}- s_{2})=  \sigma_{1} \cdot  {\hat {p}}\ \sigma_{2 }\cdot  {\hat {p}}-
\sigma_{2} \cdot  {\hat {p}}\ s_{1}\cdot  {\hat {p}} -  \sigma_1 \cdot  {\hat {p}}\  s_{2}\cdot  {\hat {p}} +  { s}_{1}\cdot { s}_{2},\label{int}
\end{equation}

which leads to
\begin{equation}
{\tilde s}_1\cdot {\tilde s}_2 = { s}_1\cdot { s}_2;
\label{int}
\end{equation}
that is, the two are equivalent. This is consistent with the study of the nucleon-nucleon interaction \cite {gino02} in which it was shown that the pseudospin transformation on two nucleons does not change the spin. However, this does not mean they are equivalent because the mixing angle between states of different pseudo-orbtial angular momentum is different than the mixing angle between states of different orbital angular momentum, which comes about through other terms involving the pseudo-orbital angular momentum operator and the orbital angular momentum operator.

Furthermore, from Eq.(\ref{gen}), the tensor interaction becomes
\begin{equation}
\sigma_1\cdot {p}\   \sigma_2\cdot  {p}\ =p^2\  [{\tilde s}_1\cdot { s}_2 + {s}_1\cdot {\tilde s}_2 +{\tilde s}_1\cdot {\tilde s}_2+{ s}_1\cdot {s}_2]. 
\label{t}
\end{equation}

That is, the tensor interaction is symmetrical in pseudospin and spin and it is an  interaction between the spin and pseudospin, which is an interesting insight.

Therefore the pseudospin interactions do not introduce additional interactions beyond the spin interactions. 

We consider now interactions involving the pseudo-orbital angular momentum. The transformation from orbital angular momentum to the pseudo-orbital angular momentum is \cite {gino02}

\begin{equation}
{\tilde L}_{i} = {\sigma_1}\cdot  {\hat p}\   {\sigma_2}\cdot  {\hat p}\ {L_i}\  {\sigma_1}\cdot  {\hat p}\   {\sigma_2}\cdot  {\hat p}
\label{oam}
\end{equation}
where the orbital angular momentum is $L_i ={ (r \times p)_i \over \hbar}$ where $r$ is the relative coordinate. The momentum does not commute with the orbital angular momentum as it does for spin and one obtains 
\begin{equation}
{\tilde L}_{i} =L_i +  2\  s_i - 2\  s\cdot  {\hat p} \  {\hat p}_i,
\label{L}
\end{equation}
where $s_i = s_{i ,1} + s_{i ,2}$ is the total spin.
The two body pseudospin-pseudo-orbit interaction becomes

\begin{equation}
{\tilde s} \cdot {\tilde L}= { -s \cdot L}+{ \sigma_{1}\cdot  {\hat p} \ \sigma_{2}\cdot  {\hat p}} + 1 - 2\  s \cdot  s ,
 \label{T}
\end{equation}

This means that this interaction can be written in terms of the two body spin-orbit interaction and the tensor interaction. On the other hand the pseudo-orbital angular momentum squared is 

\begin{equation}
{\tilde L}\cdot{\tilde L} = L\cdot L + 4\  s\cdot L - 2 +  4\  s\cdot  s - 2\
{\sigma_1}\cdot  {\hat p}\   {\sigma_2}\cdot  {\hat p}\
\label{L2}
\end{equation}

which means that the pseudo-orbital angular momentum squared can also be written in terms of the the orbital angular momentum squared, two body spin-orbit interaction and the tensor interaction. So the pseudospin and pseudo-orbital angular momentum do not introduce any new terms that are not already present with the spin and orbital angular momentum.

However, using the pseudospin and pseudo-orbital angular momentum terms  will help to elucidate the physics involved. In particular the tensor interaction breaks both orbital angular momentum and pseudo orbital angular momentum invariance whereas $s_{1}\cdot s_{2} ={\tilde s}_{1}\cdot {\tilde s}_{2}$ conserve both.  $ L\cdot L$ conserves spin symmetry and orbital angular momentum. Furthermore the addition of $s \cdot L$ produces a spin dynamical symmetry, which means that the energy levels and phase shifts will depend on the orientation of the spin and angular momentum, but the eigenfunctions will conserve spin and orbital angular momentum and the mixing angles, which measure the mixing of the orbital angular momentum in nucleon-nucleon scattering, will be zero. Likewise ${\tilde  L}\cdot {\tilde L}$ conserves pseudospin symmetry and pseudo-orbital angular momentum. Furthermore the addition of ${\tilde s} \cdot {\tilde L}$ produces a pseudospin dynamical symmetry. 
Since the tensor interaction can be written as 
\begin{equation}
\sigma_{1}\cdot  {\hat p} \ \sigma_{2}\cdot  {\hat p} = s \cdot {\tilde L} + s \cdot L - 1 + 2\ s \cdot s ,
 \label{T}
\end{equation}
the suggestion is to eliminate the tensor interaction and to write the two-nucleon interaction in momentum space in the form

\begin{eqnarray}
&~
V(p) = \nonumber \\
 &(a_{0}^{(0)}(p) + a_s^{(0)}(p)s_{1}\cdot s_{2} + a_{o}^{(0)} (p)L\cdot L+ a_{po}^{(0)}(p) {\tilde L}\cdot{\tilde L}+ a_{so}^{(0)} (p)\ s \cdot L + a_{pso}^{(0)}(p)  {\tilde s}\cdot {\tilde L}){(1-\tau_1\cdot \tau_2)\over 4}
\nonumber\\
&
+(a_{0}^{(1)}(p) + a_s^{(1)}(p) s_{1}\cdot  s_{2} + a_{o}^{(1)} (p)L\cdot L+ a_{po}^{(1)}(p) {\tilde L}\cdot{\tilde L}+ a_{so}^{(1)} (p)\ s \cdot L +a_{pso}^{(1)} (p) {\tilde s} \cdot {\tilde L}){(3+\tau_1\cdot \tau_2)\over 4},\nonumber\\
&
\end{eqnarray}
where $ \tau$ are isospin Pauli matrices and we include the possibility that the coefficients $a^{(T)}$ could depend on isospin, $T=0,1$. If the coefficients $a_{o}^{(T)},a_{so}^{(T)} $ are much larger than the coefficients $a_{po}^{(T)},a_{pso}^{(T)} $ then the orbital angular momentum will be approximately preserved whereas if the coefficients  $a_{po}^{(T)},a_{pso}^{(T)} $ are much larger than the coefficients $a_{o}^{(T)},a_{so}^{(T)} $ then pseudo-orbital angular momentum will be approximately preserved. 

In summary, we have shown that, with the introduction of the pseudospin and pseudo-orbital angular momentum operators, we have no need of the tensor interaction. Furthermore, with these new operators, the relative conservation of  orbital angular momentum and pseudo-orbital angular momentum will be transparent.

The author thanks Ionel Stetcu for discussions. This research was supported by the United States Department of
Energy under contract W-7405-ENG-36.\\

\pagebreak

\end{document}